\documentclass[reqno,12pt]{article}
\usepackage{amsmath,amsfonts,amssymb,amsthm,amstext,amscd}
\usepackage{color}
\usepackage[all]{xy}
\usepackage{hyperref}
\usepackage{epsfig}
\usepackage{graphicx}
\usepackage{tikz}
\usetikzlibrary{decorations.pathmorphing}
\usepackage{caption}
\usepackage{subcaption}
\usepackage{lipsum}
\usepackage{psfrag}
\usepackage{epstopdf}
\usepackage{comment}
\usepackage{graphicx}
\usepackage[utf8,mac]{inputenc}
\makeatletter \@addtoreset{equation}{section}

\makeatletter\renewcommand\section{\@startsection {section}{1}{\z@}%
                                                                                                                                                                                                                                                                                                                                                                                                                                                                                                                                                                                                                                                                                                                                                                                                                                                                                                                                                                                                                                                                                                                                                                                                                                                                                                                                                                                                                                                                                                                                                                                                                                                                                                                                                                                                                                                                                                   {-3.5ex \@plus -1ex \@minus -.2ex}%nn
                                                                                                                                                                                                                                                                                                                                                                                                                                                                                                                                                                                                                                                                                                                                                                                                                                                                                                                                                                                                                                                                                                                                                                                                                                                                                                                                                                                                                                                                                                                                                                                                                                                                                                                                                                                                                                                                                                   {2.3ex \@plus.2ex}%
                                                                                                                                                                                                                                                                                                                                                                                                                                                                                                                                                                                                                                                                                                                                                                                                                                                                                                                                                                                                                                                                                                                                                                                                                                                                                                                                                                                                                                                                                                                                                                                                                                                                                                                                                                                                                                                                                                   {\normalfont\large\bfseries}}
\renewcommand\subsection{\@startsection{subsection}{2}{\z@}%
                                                                                                                                                                                                                                                                                                                                                                                                                                                                                                                                                                                                                                                                                                                                                                                                                                                                                                                                                                                                                                                                                                                                                                                                                                                                                                                                                                                                                                                                                                                                                                                                                                                                                                                                                                                                                                                                                                                                                                                                                                                     {-3.25ex\@plus -1ex \@minus -.2ex}%
                                                                                                                                                                                                                                                                                                                                                                                                                                                                                                                                                                                                                                                                                                                                                                                                                                                                                                                                                                                                                                                                                                                                                                                                                                                                                                                                                                                                                                                                                                                                                                                                                                                                                                                                                                                                                                                                                                                                                                                                                                                     {1.5ex \@plus .2ex}%
                                                                                                                                                                                                                                                                                                                                                                                                                                                                                                                                                                                                                                                                                                                                                                                                                                                                                                                                                                                                                                                                                                                                                                                                                                                                                                                                                                                                                                                                                                                                                                                                                                                                                                                                                                                                                                                                                                                                                                                                                                                     {\normalfont\bfseries}}

\newcommand{\be}{\begin{equation}}
\newcommand{\ee}{\end{equation}}
\newcommand{\bea}{\begin{eqnarray}}
\newcommand{\eea}{\end{eqnarray}}
\newcommand{\bse}{\begin{subequations}}
\newcommand{\ese}{\end{subequations}}
\newcommand{\beqa}{\begin{eqnarray}}
\newcommand{\eeqa}{\end{eqnarray}}
\newcommand{\beqar}{\begin{eqnarray*}}
\newcommand{\eeqar}{\end{eqnarray*}}
\newcommand{\bi}{\begin{itemize}}
\newcommand{\ei}{\end{itemize}}
\newcommand{\ba}{\begin{array}}
\newcommand{\ea}{\end{array}}
\newcommand{\bc}{\begin{center}}
\newcommand{\ec}{\end{center}}

\def\l{\lambda}

\newcommand{\p}{\partial}

\newcommand{\mn}{{\mu\nu}}

\newcommand{\de}{\delta}

\newcommand{\eps}{\epsilon}

\newcommand{\bomega}{{\boldsymbol \omega}}
\newcommand{\bTheta}{{\boldsymbol \Theta}}

\newcommand{\bL}{{\boldsymbol L}}
\newcommand{\cL}{{\mathcal L}}
\newcommand{\ve}{{\varepsilon_0}}
\newcommand{\bA}{{\boldsymbol A}}

\newcommand{\bk}{{\boldsymbol k}}
\newcommand{\bE}{{\boldsymbol E}}
\newcommand{\bB}{{\boldsymbol B}}

\newcommand{\bv}{{\boldsymbol v}}
\newcommand{\bx}{{\mathbf x}}
\newcommand{\br}{{\boldsymbol r}}
\newcommand{\rh}{{\hat {\boldsymbol r}}}
\newcommand{\bn}{{\mathbf n}}
\newcommand{\bp}{{\mathbf p}}

\begin{document}

\begin{titlepage}

%\begin{flushright}\vspace{-3cm}
%{\small
%{\tt arXiv:yymm.nnnn} \\ IPM/P-2016/nnn \\
%\today }\end{flushright}
%\vspace{0.5cm}

\begin{center}
\centerline{\vspace{0.5cm}}
\centerline{{\huge{\sc{Multipole Charge Conservation}}}} \vspace{1mm}
\centerline{{{\sc{and}}}} \vspace{1mm}
\centerline{{\Large{\sc{Implications on Electromagnetic Radiation}}}} \vspace{12mm}

\centerline{\large{\bf{ Ali Seraj\footnote{e-mail: ali\_seraj@ipm.ir}}}}

\vspace{5mm}
\normalsize
\bigskip\medskip
\textit{School of Physics, Institute for Research in Fundamental
Sciences (IPM), \\P.O.Box 19395-5531, Tehran, Iran}\\
\vspace{5mm}

\begin{abstract}
\noindent

It is shown that conserved charges associated with a specific subclass of gauge symmetries of Maxwell electrodynamics are proportional to the well known electric multipole moments. The symmetries are residual gauge transformations surviving after fixing the Lorenz gauge, and have nontrivial charge. These ``Multipole charges'' receive contributions both from the charged matter and electromagnetic fields. The former is nothing but the electric multipole moment of the source. In a stationary configuration, there is a novel equipartition relation between the two contributions. The multipole charge, while conserved, can freely interpolate between the source and the electromagnetic field, and therefore can be propagated with the radiation. Using the multipole charge conservation, we obtain infinite number of constraints over the radiation produced by the dynamics of charged matter.
%Similarly, we propose that the residual diffeomorphism in asymptotically flat gravity in four dimensions, must reproduce the multipole moments of gravitational field.

%We compute the conserved charges associated with the infinite number of residual gauge symmetries surviving the Lorenz gauge. The multipole charge gets contribution from both the charged matter fields, and also the electromagnetic field itself. The former is exactly the electric multipole moment, while the contribution of the latter obeys a specific equipartition relation. The multipole charge, while conserved, can transmitted between the source and the EM field. Hence there can be a radiation of multipole charge to the boundary of spacetime. 
\end{abstract}

%\pacs{04.65.+e,04.70.-s,11.30.-j,12.10.-g}

\end{center}
%%%%%%%%%%%%%%%%%%%%%%%%%%%%%%%%%%%%%%%%%%%%%%%%%%%%%%%%%%%%%%%%%%%%%%%%%%%%%%%%%%%%%%%%

\end{titlepage}
%%%%%%%%%%%%%%%%%%%%%%%%%%%%%%%%%%%%%%%%%%%%%%%%%%%%%%%%%%%%%%%%%%%%%%%%%%%%%%%%%%%%%%%%%%
\setcounter{footnote}{0}
\renewcommand{\baselinestretch}{1.05}  %Line spacing
%%%%%%%%%%%%%%%%%%%%%%%%%%%%%%%%%%%%%%%%%%%%%%%%%%%%%%%%%%%%%%%%%%%%%%%%%%%%%%%%

\addtocontents{toc}{\protect\setcounter{tocdepth}{1}}
\tableofcontents

%%%%%%%%%%%%%%%%%%%%%%%%%%%%%%%%%%%%%%%%%%%%%%%%%%%%%%%%%%%%%%%%%%%%%%%%%%%%%%%%%%%%%%%%%%%
%\newpage

%%%%%%%%%%%%%%%%%%%%%%%%%%%%%%%%%%%%%%%%%%%%%%%%%%%%%%%
\section{Introduction}\label{sec-intro}
%%%%%%%%%%%%%%%%%%%%%%%%%%%%%%%%%%%%%%%%%%%%%%%%%%%%%%%
In her seminal paper, E. Noether established a profound link between the symmetries and constants of motion in the action formulation of particle or field theories \cite{Noether:1918zz}. This can also be rephrased in the Hamiltonian description where a symmetry can be associated with a function over the phase space which commutes with the Hamiltonian of the system and also generates the symmetry transformation through the Poisson bracket. These results continue to hold in the quantum theory if the symmetry is anomaly free.

Dividing symmetries of a theory into global or local, the above theorem is usually supposed to be restricted to the former. Local (gauge) symmetries, on the other hand, are considered in the Noether's second theorem (also discussed in \cite{Noether:1918zz}) which  states that the existence of local symmetries, implies a set of constraints for the theory, usually known as Bianchi identities. 
 
However, simultaneous implementation of both of the above Noether theorems leads to the association of a conserved charge to a local (gauge) symmetry as well \cite{Barnich:2001jy,Iyer:1994ys,Compere:2009dp}. The key result in this case, is that the charges can be formulated as surface (codimension 2) integrals, instead of volume (codimension 1) integrals. Accordingly, if the fields drop fast enough near the boundary, the charges associated with local symmetries would vanish. While this is usually presumed in quantum field theory, it is not the case in many examples of physical interest. A more relaxed boundary condition on gauge fields, can make the surface integral charges nonvanishing. However, one should make sure that such relaxation does not lead to divergent charges. The ``large gauge transformations'' allowed by the relaxed boundary conditions having nonvanishing charges, form a closed algebra called the asymptotic symmetry algebra.

On the other hand, one may put the boundary conditions on gauge invariant quantities like the field strength, instead of gauge field itself. This can be more physical, since the observable quantities are gauge invariant. Such boundary condition impose no restriction on the allowed gauge transformations. However, still a subclass of gauge transformations can be singled out by choosing a gauge condition instead of a boundary condition. While the gauge fixing condition kills most of the gauge redundancies, it allows for \textit{residual gauge transformations} respecting a given gauge condition. Such viewpoint was stressed in \cite{Sheikh-Jabbari:2016lzm,Avery:2015rga,Mirbabayi:2016xvc,Compere:2015knw}. In this paper, we will follow this approach and show that a subclass of residual gauge transformations can be associated with nontrivial conserved charges. In the context of Maxwell electrodynamics, we will show that such conserved quantities have a very nice interpretation in terms of electric multipole moments.

Multipole moments in Electrodynamics are obviously not conserved. For example a point charge located at origin of space has only monopole moment, while if it starts to leave the origin, it will obtain dipole and higher moments. However, we will show that if a ``soft multipole charge'' is attributed to the electromagnetic field, the total multipole charge composed of hard and soft pieces will be a conserved quantity. Interestingly, it turns out that this charge is nothing but the conserved charge associated with residual symmetries of Electrodynamics. The multipole charge can freely interpolate between the charged matter and the electromagnetic field. 

The organization of the paper is as follows. In section \ref{sec-symmetries and conservation} we derive in a systematic way, the conserved charges associated with nontrivial gauge symmetries, using the covariant phase space method. Those who are not interested in the details of the derivation can easily jump to \eqref{charges-Maxwell} and \eqref{current-J-lambda}\footnote{Although in Maxwell theory, these results can also be obtained by the usual Noether's approach, but in general the Hamiltinian approaches are preferred since e.g. in gravity the Noether charge is only a part of the correct charge \cite{Iyer:1994ys,Wald:1993nt}.}. In section \ref{sec-residual symmetries Maxwell} we determine the residual symmetries as the physical subset of $U(1)$ gauge transformations of Maxwell theory. Then in section \ref{sec-Stationary} we compute -in an electrostatic configuration- the charges associated with these symmetries and show our main result relating the charges and electric multipole moments. In section \ref{sec-Electrodynamics} we discuss electrodynamics and  show how the above conservation laws put constraints on the radiation. We conclude in section \ref{sec-discussion}. An appendix is devoted to the Poisson bracket of charges over the covariant phase space.
%%%%%%%%%%%%%%%%%%%%%%%%%%%%%%%%%%%%%%%%%%%%%%%%%%%%%%%
\section{Gauge symmetries and conservation laws}\label{sec-symmetries and conservation}
%%%%%%%%%%%%%%%%%%%%%%%%%%%%%%%%%%%%%%%%%%%%%%%%%%%%%%%
We consider the theory of Maxwell electrodynamics sourced by an arbitrary charged matter field. We choose the natural units in which $\ve=\mu_0=c=1$ and the Largrangian takes the form
\begin{align}\label{Lagrangian ED}
\cL&=-\dfrac{1}{4}F_\mn F^\mn - j^\mu A_\mu +\cL_{matter},
\end{align}
where $F_\mn=\p_{[\mu}A_{\nu]}$ is the field strength, and the current $j^\mu$ must be conserved
\begin{align}\label{J-conservation}
\p_\mu j^\mu&=0\,.
\end{align}
Variation with respect to $A_\mu$ leads to the Maxwell field equations
\begin{align}\label{Maxwell eqs}
\p_\mu F^\mn &=j^\nu \,.
\end{align}
In the next section, we give a systematic approach to compute the charges associated with gauge symmetries.
\subsection{Charges in the covariant phase space }\label{theory of charges}
In order to be able to study the conservation laws associated with gauge symmetries, one can use the Hamiltonian formulation of gauge theories \cite{Arnowitt:1962hi,Brown:1986nw,Henneaux:1992ig,Blagojevic:2002du} which is well established. However, this has the drawback that it breaks the manifest covariance of the theory, and potentially leads to cumbersome expressions. Instead, one can use a pretty mathematical construction called the ``covariant phase space'' to study gauge symmetries and associated conserved charges \cite{Lee:1990nz,Crnkovic:1986ex,Wald:1993nt,Iyer:1994ys,Barnich:2001jy,Compere:2007az} (see also \cite{Seraj:2016cym} for a review). This is the setup we use in this paper. 

To start, one needs to define a symplectic form on the space of field configurations. The symplectic form can then be used to define a Poisson bracket between functionals (of dynamical fields). Moreover, one can associate a Hamiltonian to each gauge symmetry, which generates that gauge transformation through the Poisson bracket. The on-shell value of the Hamiltonian will define the charge of that gauge symmetry. 

According to the action principle, the on-shell variation of the Lagrangian is by construction a total derivative
\begin{align}
\de\bL&\approx d\bTheta(\de\psi)\,,
\end{align}
where $\bL$ is the Lagrangian as a top form, and $\psi$ stands collectively for all dynamical fields in the theory (in our case the gauge field $A_\mu$ and the matter field $\phi$). Taking another antisymmetric variation of $\bTheta$ defines the symplectic current $\bomega$ (as a $d-1$ form of spacetime and a two form over the phase space)
\begin{align}
\bomega(\psi,\de_1\psi,\de_2\psi)&=\de_1\bTheta(\de_2\psi)-\de_2\bTheta(\de_1\psi)\,.
\end{align}
The pre-symplectic form $\Omega(\psi,\de_1\psi,\de_2\psi)$ is defined through the symplectic current $\bomega$
\begin{align}\label{symplectic form def}
\Omega(\psi,\de_1\psi,\de_2\psi)&=\int_\Sigma \bomega(\psi,\de_1\psi,\de_2\psi)\,,
\end{align}
over a spacelike hypersurface $\Sigma$ in spacetime.

A gauge theory involves local symmetry transformations of the form $\psi\to \psi+\de_\l \psi$ where $\l(x)$ is a local function (or tensor) that parametrizes the gauge transformation. 

The Hamiltonian associated to a symmetry transformation $\psi\to \psi+\de_\l \psi$ (either local or global) is then defined through
\begin{align}\label{var H def}
\de H_\l &=\Omega(\psi,\de\psi,\de_\l\psi)\,.
\end{align}
It is proved \cite{Barnich:2001jy,Iyer:1994ys} that for a gauge transformation in a gauge invariant theory one has 
\begin{align}\label{theorem-charges}
\bomega(\psi,\de\psi,\de_\l\psi)&=d\,\bk_\l (\psi,\de\psi)\,,
\end{align}
that is the symplectic current contracted with a gauge transformation is necessarily an exact form. Accordingly 
\begin{align}\label{def-Hamiltonian}
\de H_\l &=\oint_{\p\Sigma}\bk_\l (\psi,\de\psi)\,.
\end{align}
Therefore one finds the important result that the conserved charge associated with a gauge symmetry is given by a co-dimension 2 integral. Meanwhile the conserved charge of a global symmetry is given by a volume integral. This explains why the electric charge is given by the Gauss' surface integral. On the other hand it shows why energy and angular momenta are given by volume integrals in Special Relativity (where Lorentz symmetries are global) while in General Relativity, where diffeomorphisms are local symmetries, similar quantities are given by surface integrals.

Now let us compute the charges corresponding to gauge symmetries of Maxwell electrodynamics. For explicit computation, let us take the  Lagrangian of scalar QED
\begin{align}\label{scalar-QED}
\cL&=-\dfrac{1}{4}F_\mn F^\mn+ D_\mu\phi (D^\mu\phi)^*\,,
\end{align}
where $D_\mu=\p_\mu+i e A_\mu$ and its current is given by $j^\mu=i\,e\,\phi^* (D^\mu \phi)+c.c$. As we will see, the result is independent of the specific form of the matter field, and hence the rest of the paper is general for any type of electrically charged matter field. 

The theory of scalar QED is invariant under the transformations
\begin{align}\label{gauge transformations}
\de_\l A_\mu(x) =\p_\mu \l(x),\qquad \de_\l \phi(x)=ie\l(x) \,\phi(x)\,,
\end{align}
To compute the charges, let us define the dual quantities
\begin{align}
\bTheta&=\star(\theta_\mu dx^\mu),\qquad \bomega=\star(\omega_\mu dx^\mu)\,.
\end{align}
It can be checked that 
\begin{align}
\theta^\mu(\de\psi)&= F^\mn \de A_\nu +\big[(D^\mu\phi)^* \de \phi +c.c \;\big]\,,
\end{align}
and hence
\begin{align}\label{symplectic form EM}
\omega^\mu(\psi,\de_1\psi,\de_2\psi) &=\de_1 F^{\mn}\de_2 A_\nu+(\de_1(D^\mu\phi)\de_2\phi^*+c.c)-(1\leftrightarrow 2)\,.
\end{align}
To compute the charges, we need to compute $\omega^\mu(\psi,\de\psi,\de_\l\psi)$. Using gauge transformations \eqref{gauge transformations}, we arrive at 
\begin{align}\label{omega-s-QED}
\nonumber\omega^\mu(\psi,\de\psi,\de_\l\psi)&=\de F^\mn \p_\nu\l +ie \l \big(\de(D^\mu\phi)^*\phi+(D^\mu\phi)^* \de \phi\big)+c.c.\\
&=\de F^\mn \p_\nu\l +\l \de j^\mu\,.
\end{align}
Using the Maxwell equations for the linearized perturbations, we obtain the following simple form for the symplectic current
\begin{align}\label{symplectic current}
\omega^\mu(\psi,\de\psi,\de_\l\psi)&=\p_\nu (\l \de F^\mn),\qquad \text{on-shell}\,,
\end{align}
which confirms the general theorem \eqref{theorem-charges}. Accordingly the charges are defined 
\begin{align}
\de Q_\l&=\oint_{\p\Sigma} d\Sigma_\mn\, \de F^\mn \l\,,
\end{align}
and integrating over variations, one finds the finite charges 
\begin{align}\label{charges-Maxwell}
\boxed{Q_\l=\oint_{\p\Sigma} d\Sigma_\mn\, F^\mn \l(x)\,.}
\end{align}
We stress again that the charges are written only in terms of the gauge field and does not have explicit dependence on matter fields and from now on we forget the Lagrangian \eqref{scalar-QED} and work in the general case \eqref{Lagrangian ED}.
\subsection{Noether Current}
One can simply integrate over the variation in the symplectic current \eqref{symplectic current} to arrive at the Noether current 
\begin{align}\label{current-J-lambda}
J^\mu_\l&\equiv \p_\nu\big(F^\mn \l(x)\big)\,,
\end{align}
which is conserved by the antisymmetry of field strength
\begin{align}\label{conservation of currents}
\p_\mu J_\l^\mu=0\,.
\end{align}
Accordingly, the charges  \eqref{charges-Maxwell} are locally conserved in the sense that the associated current satisfies the continuity equation \eqref{conservation of currents}. We will explore the physical significance of this conservation law further in section \ref{sec-Electrodynamics}. To get more insight about $J_\l$, expand \eqref{current-J-lambda}
\begin{align}\label{current-s-h}
J^\mu_\l&=-\l(x)j^\mu+F^\mn\p_\nu\l(x)\,,
\end{align}
where we have used Maxwell equations \eqref{Maxwell eqs}. We call the first and second term ``hard'' and ``soft'' respectively. Similarly, the conserved charges can be decomposed into hard and soft pieces 
\begin{align}
Q_\l&=\int_\Sigma d\Sigma_\mu J^\mu_\l=Q_\l^{(h)}+Q_\l^{(s)}\,,
\end{align}
where
\begin{align}\label{hard and soft charges}
Q_\l^{(h)}&=-\int_\Sigma d\Sigma_\mu \l(x){j^\mu},\qquad Q_\l^{(s)}=\int_\Sigma d\Sigma_\mu F^\mn\p_\nu\l(x)\,.
\end{align}
The hard piece gives the contribution of matter source to the charge, while the soft piece gives the contribution of electromagnetic field to the charge. 

In this paper, we consider the four dimensional flat spacetime with the metric 
\begin{align}\label{metric-flat}
ds^2&=-dt^2+dr^2+r^2d\Omega^2\,,
\end{align}
though we expect that the arguments can be generalized to asymptotically flat spacetimes without much effort.
If we take the hypersurface $\Sigma$ to be the $t=const$ surface, we can use the identifications $j^\mu=(\rho,\boldsymbol{j})$, $F^{0i}=-\bE^i$ and $F_{ij}=-\eps_{ijk}\bB^k$, where $\bE,\bB$ are the spatial electric and magnetic fields respectively. Accordingly, the expressions for the charges can be simplified to 
\begin{align}\label{charge-flat}
\begin{split}
Q_\l&=-\oint_S d\vec{a} \cdot\bE \; \lambda\,,\\
Q_\l^{(h)}=-\int d^3x \l(x)&\rho,\qquad Q_\l^{(s)}=-\int d^3x  \bE\cdot\nabla \l(x)\,,
\end{split}
\end{align}
where $S$ can be chosen as a sphere of constant raduis $R\to \infty$. Note that throughout this paper $\nabla$ without explicit latin index refers to the three dimensional spatial gradient.
\subsection{Residual gauge symmetries and asymptotic symmetries}
Existence of gauge symmetries in a gauge theory provides a covariant description of the theory, at the cost of bringing in an infinite redundancy in the system. This redundancy is then removed through ``gauge fixing''. However, a specific class of gauge symmetries may survive this gauge fixing which we call \textit{residual gauge symmetries}. It turns out that a subset of residual symmetries, can be ``large'' near the boundary. It is argued in many different ways that ``large gauge symmetries'' can play important physical role in different theories. Most famously, in the context of gravity, large gauge transformations provide basic understanding of holography \cite{Brown:1986nw,Skenderis:2002wp}, microscopic counting of black hole entropy \cite{Strominger:1997eq,Carlip:1998wz,Sheikh-Jabbari:2016lzm}, and even an identification of black hole microstates \cite{Afshar:2016uax}. In QED and gravity, they are recently used to prove Weinberg's soft theorems  \cite{Strominger:2013lka,He:2014cra,He:2014laa,Kapec:2015ena,Mirbabayi:2016xvc}. Also large gauge transformations are used to describe the so called ``edge states'' in quantum Hall effect \cite{Balachandran:1991dw,Balachandran:1994up}. The role of such symmetries in theories with Weyl scaling is still unclear \cite{Jackiw:2015aoc}.

We should make a comparison here between the notions of residual gauge symmetry and the more familiar \textit{asymptotic symmetry}. An asymptotic symmetry is defined through a consistent boundary condition on the gauge fields. Boundary conditions rule out too large gauge transformations which break the BCs. The remaining ones, are called nontrivial (trivial) if the associated charge is nonvanishing (vanishing). Asymptotic symmetries are defined as the quotient of nontrivial modulo trivial gauge transformations. Through the Lie bracket, they form the asymptotic symmetry algebra. While many intriguing results have been obtained from this approach, it has the drawback that the boundary conditions must be imposed on gauge fields (like $A_\mu$) and not on gauge invariant quantities (like $F_\mn$). However, the physical significance of such boundary conditions is not clear, since local physics involves gauge invariant quantities. The situation is different in gravity, since in that case such conditions can be interpreted as the choice of ``observers'' at infinity.

On the other hand, if one fixes boundary conditions on gauge invariant quantities, no gauge transformation is ruled out by the boundary conditions. Then the ``physical'' gauge symmetries are the nontrivial residual symmetry transformations that respect the gauge condition. This has the advantage that the form of residual symmetries are determined all-over the bulk, and not only at the boundary. Unlike the asymptotic symmetries, the radial dependence of these symmetries can be completely different from one to another. This has interesting implications that we will discuss in our problem. However, this approach can have its own drawbacks. The form of residual symmetries depend on the choice of gauge condition which cannot be singled out by physical considerations. Still some gauge conditions are favored e.g. by requiring causality in the propagation of gauge field. We expect that the results must be eventually independent of the gauge condition, however a general proof of such claim remains as an open problem (more comments can be found in \cite{Avery:2015rga}).

%%%%%%%%%%%%%%%%%%%%%%%%%%%%%%%%%%%%%%%
\section{Residual gauge symmetries of Maxwell theory}\label{sec-residual symmetries Maxwell}
%%%%%%%%%%%%%%%%%%%%%%%%%%%%%%%%%%%%%%%
In this section, we determine the residual gauge symmetries of Maxwell theory in Lorenz gauge. Then we single out the nontrivial sector of these symmetries. 

To remove the infinite redundancy in Maxwell theory due to $U(1)$ gauge degrees of freedom, we impose the Lorenz gauge condition 
\begin{align}\label{Lorentz gauge}
\nabla_\mu A^\mu&=0\,.
\end{align}
In this gauge, Maxwell equations become wave equations $\Box A^\mu={j^\mu}$. However, there are still residual gauge transformations that respect the Lorentz gauge and therefore are not killed by the gauge fixing. They satisfy the equation
\begin{align}\label{guide eq}
\Box \l(x)&=0\,.
\end{align}
Expanding the time dependence of $\l$ in Fourier modes 
\begin{align}
\l(t,\bx)&=e^{-i\omega t} \l_\omega (\bx)\,,
\end{align}
the equation becomes the Helmholtz equation
\begin{align}
(\nabla^2+\omega^2)\l_\omega&=0\,.
\end{align}
There are two qualitatively different sets of solutions: those with $\omega\neq 0$ and those with $\omega=0$. They have drastically different behavior at large radial coordinates \footnote{Similarly the equation $(\frac{\p^2}{\p x^2}+\omega^2)f(x)=0$ has oscillatory solutions while the solutions to $\frac{\p^2}{\p x^2}f(x)=0$ grow linearly in $x$. }. The general solution to the wave equation with nonvanishing frequency is 
\begin{align}\label{Helmholtz sol}
\l_\omega(\bx)&=\sum_{\ell=0}^\infty\sum_{m=-\ell}^{\ell}\left(A_{\ell,m} h^{(+)}_\ell(kr) +B_{\ell,m} \,h^{(-)}_\ell(kr)\right)\;Y^*_{\ell,m}(\theta,\varphi)\,,
\end{align}
in which $h^{(+)},h^{(-)}$ are respectively the spherical Hankel functions of first and second kind, which express outgoing and ingoing waves, respectively. The asymptotic expansion of these functions are given by 
\begin{align}
h^{(\pm)}_\ell(x)&=\dfrac{(\mp i)^{\ell+1}\,e^{\pm ix}}{x}+\mathcal{O}(\dfrac{1}{x^2})\,.
\end{align}
As we will see in the next section, any solution of the form of \eqref{Helmholtz sol} is trivial in the sense that the corresponding charge is vanishing. Meanwhile, the story is different for the vanishing frequency modes, which satisfy the Laplace equation $\nabla^2\l(\bx)=0$ whose solutions are given by 
\begin{align}\label{residual symmetries}
\l(\bx)&=-\sum_{\ell=0}^\infty\sum_{m=-\ell}^{\ell}(c^+_{\ell,m} \,r^\ell +c^-_{\ell,m} \,r^{-(\ell+1)})\;Y^*_{\ell,m}(\theta,\varphi)\\
&\equiv -(c^+_{\ell,m} \lambda^+_{\ell,m}+c^-_{\ell,m} \lambda^-_{\ell,m})\,.
\end{align}
The minus sign is chosen to cancel the minus signs appearing in \eqref{charge-flat}. We will see shortly that while the negative power modes are again trivial, those with positive power (those with $c^+_{\ell,m}\neq 0$) have a non-vanishing charge with interesting physical interpretation.
\subsection{The nontrivial sector }
Using \eqref{charge-flat}, we can associate a charge to each of the above residual symmetries
\begin{align}\label{charge-pm}
Q^\pm_{\ell,m}&=-\oint d\vec{a} \cdot\bE \; \lambda^\pm_{\ell,m}\,,
\end{align}
where the integral is taken over a sphere of constant raduis $R\to \infty$. To compute this integral, we need to specify the asymptotic behavior of $\hat{r}\cdot \bE=F^{r0}$. To this end, we note that the scalar $\br\cdot\bE$ solves the same equation as \eqref{guide eq} whose solutions are given by \eqref{Helmholtz sol} which falloff like $\mathcal{O}(1/r)$. Therefore assuming that the sources are localized, the reasonable boundary condition is 
\begin{align}\label{BC}
\hat{r}\cdot \bE\sim \mathcal{O}(\dfrac{1}{r^2})\,.
\end{align}
This can be considered as a Neumann boundary condition which allows most of physically interesting situations including radiating systems. \footnote{Note that in Electromagnetic radiation, only the transverse components of the electric and magnetic field fall off as $1/r$.} It is important to note that this boundary condition does not impose any restriction over gauge transformations, since $F_\mn$ is gauge invariant. Therefore all of $\l^\pm_{\ell,m}$ transformations are allowed. This makes our approach different with the usual asymptotic symmetry group analysis.

As was promised, we can now show that only the ``soft part'', i.e. the zero frequency subset of the residual symmetries lead to nontrivial charges. It is enough to use \eqref{BC} and \eqref{residual symmetries} in \eqref{charge-pm} to arrive at 
\begin{align}
Q_{\l_\omega}&\sim \mathcal{O}(\dfrac{1}{r})\to 0\,.
\end{align}
Similar reasoning implies that the charge of $\l^{(-)}_{\ell,m}$ in \eqref{residual symmetries} is vanishing. That is,  $\l^{(-)}_{\ell,m}$ and $\l_\omega$ are pure gauge transformations, since their charge cannot be used to label different configurations of the phase space. In \cite{Mirbabayi:2016xvc}, the same conclusion was arrived at, using another argument based on the notion of adiabatic modes \cite{Weinberg:2003sw}. \footnote{Also it is interesting to note that, in vacuum, the above pure gauge transformations can be removed further by the additional gauge fixing condition $A_0=0$ without affecting the physical residual symmetries. But this is not what we will do here.}

On the other hand, $\l^{(+)}_{\ell,m}$ transformations grow badly in large radius and one may expect that the corresponding charges diverge. However, as we will see, the existence of spherical harmonics kills all divergent terms and leads to well defined, physically meaningful charges. The zero mode $\l^{(+)}_{0,0}=\frac{1}{4\pi}$ corresponds to 
\begin{align}
Q^+_{0,0}&=\frac{1}{\sqrt{4\pi}}\oint d\vec{a} \cdot\bE\,.
\end{align}
This implies that total electric charge is the charge of constant gauge transformation of Electromagnetic theory. Note that this gauge transformation is special, since it leaves the gauge field intact, i.e. $A\to A+d\l=A$. We will come back to this point later in section \ref{sec-symplectic symmetries}. 

Therefore the conservation of electric charge is a direct consequence of gauge invariance of the theory. However, this is not the whole information that can be inferred from the gauge invariance. This is what we show in the next sections by computing the charge corresponding to $\l^+_{\ell,m}$
\begin{align}\label{charge-plus}
Q_{\ell,m}&\equiv\oint d\vec{a} \cdot\bE\, r^{\ell} Y^*_{\ell,m}\,.
\end{align}
Note that hereafter, we drop the plus index of $Q^+_{\ell,m}$, since the minus sector is trivial. 

Before closing this section, we should briefly mention about the commutator of charges over the phase space. We show in the appendix that the Poisson bracket between the Hamiltonian generators $H_{\l}$ of symmetry transformation parametrized by $\l$ is given by (for a more detailed discussion, see \cite{Seraj:2016cym})
\begin{align}
\{H_{\l_1},H_{\l_2}\}&=\int_\Sigma \bomega(\psi,\de_{\l_1}\psi,\de_{\l_2}\psi)\,.
\end{align}
Using the symplectic form if Maxwell theory \eqref{symplectic form EM}, and the transformation rules \eqref{gauge transformations}, we find that the Poisson bracket of charges vanish. 
\begin{align}
\{H_{\l_1},H_{\l_2}\}&=0\,.
\end{align}
The charges, being the on-shell value of the Hamiltonian generators, obey the same algebra. This agrees with the general theorem that the Poisson bracket of charges is a central extension of the Lie algebra of symmetries up to possibly a central extension \cite{Brown:1986ed,Barnich:2001jy,Compere:2007az}
\begin{align}
\{H_{\l_1},H_{\l_2}\}&=H_{[\l_1,\l_2]}+C(\l_1,\l_2)\,.
\end{align}
Here the Lie algebra of $U(1)$ gauge symmetries is trivial and no central extension arise at the level of charges.
%%%%%%%%%%%%%%%%%%%%%%%%%%%%%%%%%%%%%%%%%%%%%%%%%%%%%%%
\section{Stationary configurations}\label{sec-Stationary}
%%%%%%%%%%%%%%%%%%%%%%%%%%%%%%%%%%%%%%%%%%%%%%%%%%%%%%%
In this section, we compute the charges corresponding to nontrivial residual symmetries in a stationary solution of Maxwell theory. This will provide us a physical interpretation of these symmetries.
\subsection{Electrostatics}
In this case, $\bB=0$ and we can write $\bE=-\nabla\Phi$. The potential obeys the Laplace equation whose solution can be expanded as
\begin{align}\label{Multipole expansion}
\Phi(\bx)&=\dfrac{1}{4\pi}\sum_{\ell,m}\dfrac{q_{\ell,m}}{r^{\ell+1}}\dfrac{4\pi}{2\ell+1}Y_{\ell,m}(\theta,\varphi)\,.
\end{align}
The coefficients $q_{\ell,m}$ are called  the ``electric multipole moments'' which are determined by the distribution of charged matter as
\begin{align}
q_{\ell,m}&=\int d^3x \,\rho(\bx)\, r^\ell Y^*_{\ell,m}\,.
\end{align}
Hence an electrostatic configuration is completely determined, given the multipole moments. Note also that higher order moments fall off more and more rapidly in large distances. Now let's compute the charges \eqref{charge-plus} an electrostatic configuration of the above form,
\begin{align}
Q_{\ell,m}&=-\oint d\vec{a} \cdot{\nabla}\Phi\, r^{\ell} Y^*_{\ell,m}\,.
\end{align}
Using \eqref{Multipole expansion} and the fact that the integral is taken over a sphere at large $R$ we have
\begin{align}
\nonumber Q_{\ell,m}&=-\oint R^2\, d\Omega \;\p_r\Phi\; R^{\ell} Y^*_{\ell,m}\\
&=\dfrac{1}{4\pi}\sum_{\ell',m'}(\ell'+1)\dfrac{q_{\ell',m'}}{R^{\ell'+2}}\dfrac{4\pi}{2\ell'+1}R^{\ell+2}\oint d\Omega Y_{\ell',m'}(\theta,\varphi)\,  Y^*_{\ell,m}\,.
\end{align}
Given the orthogonality of spherical harmonics $\oint d\Omega Y_{\ell',m'}(\theta,\varphi)\,  Y^*_{\ell,m}=\delta_{\ell,\ell'}\,\de_{m,m'}$, we conclude our main result
\begin{align}\label{multipole charges-static}
\boxed{Q_{\ell,m}=\dfrac{\ell+1}{2\ell+1}q_{\ell,m}\,.}
\end{align}
This result implies that the charges associated with physical residual gauge symmetries are proportional to the \textit{electric multipole moments}\footnote{In SI units, the above would read $Q_{\ell,m}=\frac{\ell+1}{2\ell+1}\frac{q_{\ell,m}}{\ve}$ where $\ve$ is the  vacuum permittivity. }. This gives the classical interpretation of residual symmetries of Electrodynamics. Accordingly, we call $Q_{\ell,m}$ the  \textit{multipole charge}. %In terms of multipole charges, the potential can be written as 
%\begin{align}
%\Phi(\bx)&=\sum_{\ell,m}\dfrac{Q_{\ell,m}}{r^{\ell+1}}(\ell+1)Y_{\ell,m}(\theta,\varphi)\,.
%\end{align}

The factor $\frac{\ell+1}{2\ell+1}$ is also important. For a better understanding of this factor, let's compute the hard and soft contributions to the charge $Q_{\ell,m}$ as defined in \eqref{hard and soft charges},
\begin{align}
Q_{\ell,m}^{(h)}&=\int d^3x \,\rho \,\lambda_{\ell,m}=\int d^3x \,\rho\, r^\ell Y_{\ell,m}(\theta,\varphi)=q_{\ell,m}\,.
\end{align}
Therefore the hard piece of charge exactly reproduces the multipole moment of order $(\ell,m)$. However, there is also a contribution from the fields, i.e. the soft piece
\begin{align}\label{soft piece}
Q_{\ell,m}^{(s)}&=Q_{\ell,m}-Q_{\ell,m}^{(h)}=-\dfrac{\ell}{2\ell+1}\,q_{\ell,m}\,.
\end{align}

\subsubsection{Screening effect and equi-partition relation}
As we mentioned, the hard piece is the contribution from matter fields to the charges $Q_{\ell,m}$, while the soft piece is the contribution from the surrounding electromagnetic field. For the case $\ell=0$, which corresponds to the total electric charge, we see from \eqref{soft piece} that electromagnectic field does not carry any electric charge. This is what we expect from a $U(1)$ gauge theory. However, it \textit{does} carry higher multipole charges. 

The minus sign in \eqref{soft piece} means that in the equilibrium, there is an \textit{screening effect} from the fields reducing the effective multipole charge $Q_{\ell,m}$ compared to the bare (hard) multipole charge $Q^{(h)}_{\ell,m}$. The above result can also be written in the suggestive form
\begin{align}
Q_{\ell,m}^{(s)}&=-\dfrac{\ell}{2\ell+1}Q^{(h)}_{\ell,m}\,,
\end{align}
which resembles a special ``equipartition'' relation between the soft and hard pieces of the multipole charge. Note that for $\ell\gg 1$ this approaches $-\frac{1}{2}$. It is tempting to find a deeper understanding of this equipartition relation of multipole charges in equilibrium between electromagnetic field and matter source.

\subsubsection{Symplectic symmetries}\label{sec-symplectic symmetries}
In \cite{Compere:2014cna,Compere:2015bca,Compere:2015mza} the notion of \textit{symplectic symmetry} was defined by the condition that the symplectic current \eqref{symplectic current} vanishes locally outside sources. Accordingly, using \eqref{theorem-charges},\eqref{def-Hamiltonian} it can be easily shown that the charges can be computed at any surface sorrounding the sources, not only at the boundary\cite{Hajian:2015xlp}.

Among the multipole charges, only $Q_{0,0}$ corresponding to the total electric charge is precisely symplectic, since according to \eqref{current-s-h} the current vanishes identically outside sources. 

For other multipole charges, the situation is different. In the electrostatic case, it can be checked that although they are not symplectic in the strict sense, but they can still be computed at any closed sphere \textit{containing} the source. The reason is that, while the symplectic current \eqref{symplectic current} or its finite version \eqref{current-J-lambda}, is not vanishing locally, its volume integral over regions free of charged matter vanishes, i.e. 
\begin{align}
Q\Big\vert_{S_2}-Q\Big\vert_{S_1}&=\int_{\Sigma_{12}}d\Sigma_\mu J^\mu_\l =0\,,
\end{align}
where $S_{1,2}$ are two spheres with different radii, both containing the source, and $\Sigma_{12}$ is the volume enclosed between the two spheres. Therefore the soft piece of multipole charge only gets nontrivial contribution within the charge \textit{horizon} \cite{Compere:2016hzt}, i.e. the smallest sphere containing sources.

However, this does not continue to hold when radiation enters in the game, which carries nontrivial multipole charges except $Q_{0,0}$. Therefore, symplectic symmetries appear in the nondynamical sector of the phase space in accordance with results of \cite{Compere:2015bca,Compere:2015mza}.

\subsection{Magnetostatics}\label{sec-magnetostatics}
We showed in previous section that the nontrivial gauge symmetries are associated to \textit{electric} multipole charges. These charges are blind to the magnetic field, and hence cannot uniquely fix the field. In order to overcome this deficiency, we need to define a new set of charges that correspond to magnetic multipoles\footnote{The author is grateful to Jarah Evslin, Temple He, and Shahin Sheikh-Jabbari for useful discussions on this section.}. Such charges were introduced in \cite{Strominger:2015bla} (see also \cite{Campiglia:2016efb}) through the electic-magnetic duality
\begin{align}\label{Magnetic charge}
\tilde{Q}_\l&\equiv \oint_S d\Sigma_\mn \,\varepsilon^{\mn\alpha\beta}F_{\alpha\beta}\,\l(x)=\oint_S  da\cdot\bB\,\l(x)\,,
\end{align}
where $\varepsilon^{\mn\alpha\beta}$ is the Levi-Civita symbol, and the gauge parameter $\l(x)$ is any combination of solutions of Laplace equation of the form $r^\ell Y^*_{\ell,m}$. In differential forms language where $F$ is a two form, the charges \eqref{charges-Maxwell} can be related to the three form Noether current $J_\l=d(\l \star F)$ which is conserved on-shell, while the above charge corresponds to the \textit{off-shell} conserved current 
\begin{align}
J_\l&=\mathrm{d}\,(\l F)=d\l\wedge F\,.
\end{align}
In the last equation, we have used the Bianchi identity $dF=0$. The charges can accordingly be written as volume integrals 
\begin{align}
\tilde{Q}_\l&=\int d^3x \bB\cdot\nabla\l\,.
\end{align}
We observe that this is similar to the electric result \eqref{charge-flat}, but without the hard contribution. Specifically, the monopole charge corresponding to $\l=1$ vanishes. In the following, we compute the magnetic charge \eqref{Magnetic charge} for a magnetostatic configuration. In this case, outside the source, the magnetic field can be written as $\bB=-\nabla \Phi_M$ where \cite{bronzan1971magnetic}
\begin{align}\label{Magnetic Multipole expansion}
\Phi_M(\bx)&=\dfrac{1}{4\pi}\sum_{\ell,m}\dfrac{M_{\ell,m}}{r^{\ell+1}}\dfrac{4\pi}{2\ell+1}Y_{\ell,m}(\theta,\varphi)\,.
\end{align}
The coefficients $M_{\ell,m}$ are the ``magnetic multipole moments'' given by 
\begin{align}
M_{\ell,m}&=-\dfrac{1}{\ell+1}\int d^3x \, r^\ell Y^*_{\ell,m}\,\nabla\cdot(\br\times \boldsymbol{j}).
\end{align}
Using \eqref{Magnetic Multipole expansion} in \eqref{Magnetic charge}, we find 
\begin{align}
\tilde{Q}_{\ell,m}&=\dfrac{\ell+1}{2\ell+1}\,M_{\ell,m}
\end{align}

We finish this section by noting that since Maxwell theory is linear, one can superpose arbitrary magnetostatic and electrostatic solutions to obtain a general stationary solution. The multipole charges \eqref{charge-flat} and \eqref{Magnetic charge} detect the electric and magnetic distributions respectively, and are blind to the other. We will use these results in section \ref{sec-Electrodynamics}.

%%%%%%%%%%%%%%%%%%%%%%%%%%%%%%%%%%%%%%%%%%%%%%%%%%%%%%%
\section{Electrodynamics}\label{sec-Electrodynamics}
%%%%%%%%%%%%%%%%%%%%%%%%%%%%%%%%%%%%%%%%%%%%%%%%%%%%%%%
Before studying the charges in the Electrodynamic case, let us discuss the conservation law \eqref{conservation of currents} for $\l_{\ell,m}$ symmetries in more detail. 
\subsection{Conservation}
Let us expand the continuity equation \eqref{conservation of currents}, using $j^\mu=(\rho,\boldsymbol{j})$
\begin{align}
\dfrac{d}{dt}\left(\rho\, \l+\bE\cdot\nabla\l \right)+\nabla\cdot\left(\l\, \boldsymbol{j}+\bB\times\nabla\l\right)&=0\,.
\end{align}
where $\l$ is a combination of nontrivial residual symmetries found in section \ref{sec-residual symmetries Maxwell}. Integrating this over a $t=const$ hypersurface and using the definition of hard and soft pieces of charge \eqref{hard and soft charges}, we obtain
\begin{align}\label{conservation-flux}
\dfrac{d}{dt}Q_\l&=\dfrac{d}{dt}\left(Q^{(h)}_\l+Q^{(s)}_\l\right)=-\oint da\cdot\left(\l\,\boldsymbol{j}+\bB\times\nabla\l\right)\equiv \mathcal{F}_\l\,.
\end{align}
This implies that the time rate of change of multipole charge stored in both the source and EM field equals minus the flux of multipole current at the boundary. This is the statement of the conservation of multipole charge. Note that the usual expression for multipole moment i.e. $Q^{(h)}_\l$ is not conserved individually, since the multipole charge can freely interpolate between electromagnetic field and charged matter. To see this explicitly, define ${J^\mu}^{(h)}_\l\equiv\l j^\mu$, representing only the hard piece of multipole current whose integral gives $Q^{(h)}_\l$. Then multiply  \eqref{J-conservation} by $\l$ and use the time independence of $\l$ to arrive at 
\begin{align}\label{hard-nonconservation}
\p_\mu {J^\mu}^{(h)}_\l&=\boldsymbol{j}\cdot \nabla \l\,.
\end{align}
This gives the transfer rate of multipole charge from the electromagnetic field to the source, and makes clear why the hard charge is not conserved.
%\fixme{while there is no electric multipole charge in magnetostatic, the rhs seems to be nonvanishing. What happens?}
\subsection{A preliminary example}
A simple intuition that may stop one to think of multipole moments as conserved charges is that a point charge with constant velocity has an increasing dipole moment growing with time. Before studying the real dynamical situations, let us discuss this example. While this is trivial given the fact that it can be reverted to the electrostatic case through a Lorentz transformation, it will be illuminating in some aspects.

\begin{figure}[!h]
\centering
\includegraphics[width=0.6\linewidth]{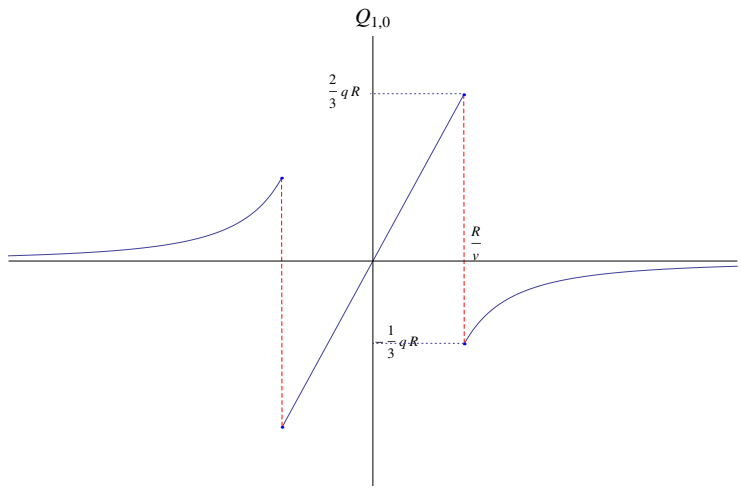}
\captionsetup{width=0.8\textwidth}
\caption{{\small The dipole charge within a sphere of radius $R$, produced by a charged particle moving with constant velocity along the $z$ direction. The discontinuities occur when the particle enters and exits the integration surface.}}
\label{fig:moving particle}
\end{figure}

Consider a particle moving with a constant velocity $v$ along the $z$ direction. Its current is given by $j^\mu=\rho \,\frac{dx^\mu}{dt}$ and 
\begin{align}
\rho&=q \,\de^3\left(\br-\br_0(t)\right)\,,
\end{align}
where $\br_0(t)= v\,t\,\hat{z}$ is the position of the point charge. The particle produces electric and magnetic fields
\begin{align}
\label{E field-particle}\bE&=\dfrac{\gamma \overline{\bE}}{\big(1+ (\gamma\,\bv\cdot\bn)^2\big)^{3/2}}\,,\\
\label{B field-particle}\bB&=\bv\times\bE\,,
\end{align}
where $\overline{\bE}=\frac{q (\br-\br_0)}{4\pi |\br-\br_0|^3}$, $\gamma=(1-v^2)^{-1/2}$ and $\bn$ is a unit radial vector from the charge's present position to the observation point. Now the dipole moment (the hard piece of dipole charge) is 
\begin{align}\label{dipole}
Q^{(h)}_{1,0}(t)&={q_{1,0}}={q v t}\,,\qquad|t|<\dfrac{R}{v},
\end{align}
which is linearly growing in time. Meanwhile, according to \eqref{hard-nonconservation}, the rate of transfer of dipole charge between the source and the field is 
\begin{align}
\int d^3x \,\boldsymbol{j}\cdot \nabla \l_{1,0}&=\int d^3x J_z= {q v}\,,
\end{align}
which is exactly the time rate of change of \eqref{dipole}. Therefore the change in dipole moment of the point charge is due to the absorption of dipole charge from the electromagnetic field. The total dipole charge is off by a factor $2/3$ as in the previous section due to the soft part stored in the field. Therefore the total charge is not constant in time. This is possible only if there is a flux of dipole charge at the boundary. To see this, assume that the integration surface is a large sphere at $r=R$. There is no flux of charged particles at the boundary when $|t| < R/v$. The flux of dipole charge in this period,  can be obtained using \eqref{B field-particle} in \eqref{conservation-flux}, leading to $\mathcal{F}_{1,0}=-\frac{2}{3}q v$ as expected. After a period $\Delta t=R/v$, the total incoming flux is $\frac{2}{3}{q R}$. At $t=R/v$, when the particle escapes the integration surface, there is a sudden outgoing flux of hard charge by the amount $+{q R}$. The difference is nothing but the soft charge remaining in the integration surface. As the particle gets farther, the soft charge within the integration surface decays as
\begin{align}
Q_{1,0}(t)&=Q_{1,0}^{(s)}(t)=-\dfrac{1}{3}{q}\dfrac{R^3}{z(t)^2}\,\qquad |t|> \dfrac{R}{v}\,.
\end{align}
These are summarized in figure \ref{fig:moving particle}.

\subsection{Infinite constraints over radiation}\label{sec-radiation constraints}
In this section, we study how the multipole charge conservation constrains the radiation generated by a dynamical charged system. Throughout this section, we assume that the dynamics takes place in a region of spacetime with compact support. This is reasonable since an eternal dynamical radiating system requires an infinite source of energy. Therefore we consider the matter configuration which is stationary in the region $|t|>T$ and  the dynamics happens in the interval $|t|<t_0$ ($t_0$ can be any given number). We first prove that the charge is conserved in time, and then show how this conservation laws impose infinite number of constraints over the radiation produced by the source. Figure \ref{fig:moving particle} is a schematic picture of the problem.
\begin{figure}[h]
\centering
\includegraphics[width=0.25\linewidth]{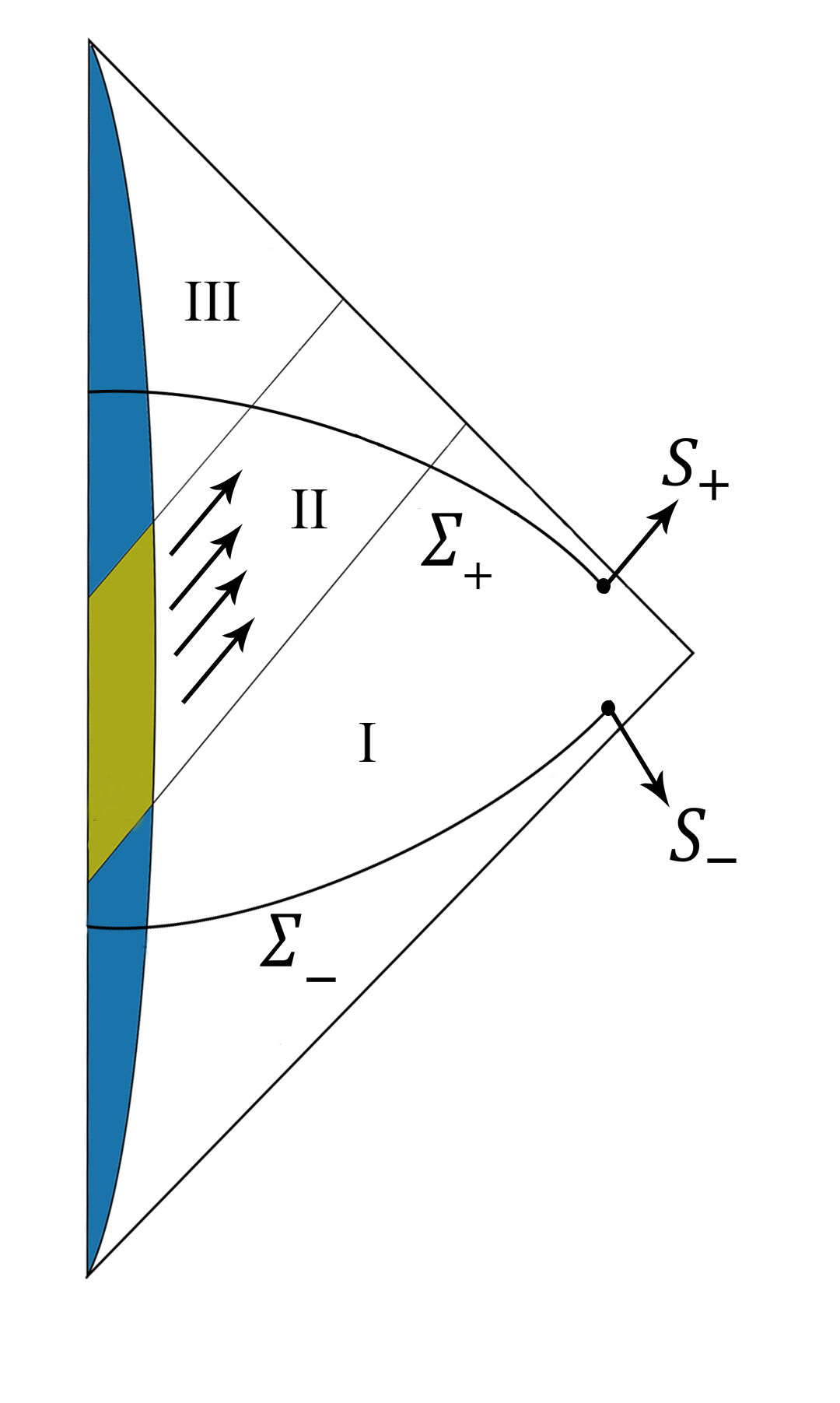}
\captionsetup{width=0.8\textwidth}
\caption{{\small A distribution of charged matter (colored), which is stationary in far past and far future (regions I,III) and radiating in the region II. The charges are computed at two time slices $\Sigma_-$ and $\Sigma_+$ before and after the radiation phase,  and are given by surface integrals over $S_\pm$ of radius $R$ where $S_\pm=\p \Sigma_\pm$. These two surfaces are connected by a timelike hypersurface $\Sigma_B$ of constant radius (not drawn). Note that both $S_\pm$ reside in region I as $R \to \infty$. }}
\label{fig:figure-2}
\end{figure}

As we mentioned before, associated with any residual symmetry $\l_{\ell,m}=r^\ell Y^*_{\ell,m}$, there is a charge $Q_{\ell,m}(t,R)$ computed at a sphere of radius $R$ at a constant time slice $t$. Further as before, we assume that $R\to \infty$. Now let us compute the charge at two different times $t=\pm T$ where $T > t_0$, that is before and after the dynamics of the source. As mentioned before, the charge at $t=\pm T$ is given by an integral over its boundary $S_\pm$. The important point is that since $R$ is taken to infinity i.e. $R\gg cT$, both $S_\pm$ fall in the region $A$ of the spacetime. Using this we can show that $Q_+=Q_-$. The reason is that
\begin{align}
Q^{(+)}_\l-Q^{(-)}_\l&=\int_{\Sigma_B}\mathcal{F}_\l\,,
\end{align}
where $\Sigma_B$ is the timelike boundary between $S_-, S_+$ and the flux $\mathcal{F}$ is given by \eqref{conservation-flux}. However, the flux is zero since no radiation can reach $\Sigma_B$ as $R\to \infty$, as it is clear from figure \ref{fig:figure-2}. Note that we have assumed that the source of radiation has compact support in space and time. Accordingly, we obtain the conservation of multipole charges in the presence of radiation
\begin{align}
Q^{(+)}_{\ell,m}&=Q^{(-)}_{\ell,m}\,.
\end{align}
Moreover, the charge at $t=-T$ is given by \eqref{multipole charges-static}, i.e.
\begin{align}
Q^{(-)}_{\ell,m}&=\dfrac{\ell+1}{2\ell+1}{q_{\ell,m}^{(-)}}\,,
\end{align}
where $q_{\ell,m}^{(-)}$ denotes the multipole moments in the stationary phase before the dynamics. Now let us compute $Q^{(+)}$ through a volume integral over $\Sigma_+$. The constant time hypersurface $\Sigma_+$ naturally divides into three regions $\Sigma_+^\mathrm{I},\Sigma_+^{\mathrm{II}},\Sigma_+^{\mathrm{III}}$ as shown in figure \ref{fig:figure-2}. Therefore
\begin{align}
Q^{(+)}_\l&=\int_{\Sigma_+}J_\l=\int_{\Sigma_+^\mathrm{I}}J_\l+\int_{\Sigma_+^\mathrm{II}}J_\l+\int_{\Sigma_+^\mathrm{III}}J_\l\,.
\end{align}
According to the discussion in section \ref{sec-symplectic symmetries}, the first term on the right hand side is zero, while the last term is 
\begin{align}
\int_{\Sigma_+^\mathrm{III}}J_{\ell,m}&=\dfrac{\ell+1}{2\ell+1}{q_{\ell,m}^{(+)}}\,.
\end{align}
Therefore we find that the total multipole charge $Q_{\ell,m}^{rad}$ carried by the radiation is 
\begin{align}\label{radiation constraints}
\boxed{Q_{\ell,m}^{(rad)}=\dfrac{\ell+1}{2\ell+1}\left(q_{\ell,m}^{(-)}-q_{\ell,m}^{(+)}\right)\,.}
\end{align}
Similarly, the same argument can be repeated for the magnetic multipole charges $\tilde{Q}_{\ell,m}$ discussed in section \ref{sec-magnetostatics}, to obtain another set of constraints over the radiation 
\begin{align}\label{radiation constraints magnetic}
\tilde{Q}_{\ell,m}^{(rad)}=\dfrac{\ell+1}{2\ell+1}\left(M_{\ell,m}^{(-)}-M_{\ell,m}^{(+)}\right)\,.
\end{align}
Given merely the initial and final stationary configuration of the matter, determined by $q_{\ell,m}^{(\pm)},\, M_{\ell,m}^{(\pm)}$, we have found infinitely many constraints over the radiation produced during the dynamical phase of the system, without solving the equations of motion. This result resembles the recent developments relating the asymptotic symmetries of QED with Weinberg soft photon theorem \cite{He:2014cra,Kapec:2015ena,Strominger:2015bla}. However, finding a precise relation is beyond the scope of this paper.

%%%%%%%%%%%%%%%%%%%%%%%%%%%%%%%%%%%%%%%%%%%%%%%%%%%%%%%
\section{Discussion}\label{sec-discussion}
In this paper, we studied the conservation laws associated with residual symmetries of Maxwell theory. We showed that among the residual gauge transformations surviving the Lorenz gauge, only those solving the Laplace equation and are growing in large radius correspond to nontrivial conserved charges. Interestingly, these charges turned out to be proportional to the multipole moments of the charged matter distribution, hence dubbed as ``multipole charges''. The multipole charge is not equal to electric multipole moment, since the electromagnetic field gives a soft contribution to the multipole charge, and this is exactly what makes the multipole charge conserved, while the multipole moment is obviously not conserved. The only exception is the electric monopole which is only stored in the charged matter. Using the electric-magnetic duality, we also defined the magnetic multipole charges proportional to to magnetic multipole moments.

Using the conservation of multipole charges, we found infinite number of constraints over the radiation produced by the charged matter, without knowing about the dynamics of the source which can be in general complicated.

This analysis can be followed in different directions that we mention in the following. While the electric multipole charges are Noether charges derived from residual gauge symmetries, a first principle derivation of magnetic charges remains as an open issue. Unlike the magnetic monopole charge (which is zero in our problem), higher multipole charges are not topological since the field is smooth everywhere outside the source.

Although the analysis in this paper was done for flat spacetime in four dimensions, we expect that similar analysis can be carried out for asymptotically flat spacetimes in arbitrary dimensions. In case of asymptotically flat black holes geometries, one should note that a part of radiation can be absorbed by the black hole. In this case, it was shown in \cite{Avery:2016zce} that the absorption rates of long wavelength radiation is determined by the conservation of energy and large gauge transformations. 

The same study may be repeated in gravity where the gravitational multipole expansion is well established \cite{Thorne:1980ru}, but to our knowledge never studied in relation with conservation laws of residual symmetries. The constraints over the gravitational radiation can be especially interesting due to the recent detection of gravitational waves from black hole mergers. 

While there is qualitative difference in radial dependence of our residual symmetries with the asymptotic symmetries  considered in \cite{Strominger:2013lka,He:2014cra,Kapec:2015ena,Conde:2016csj}, we still expect that there is a close link between equation  \eqref{radiation constraints} and their results. The reason is that there is a one to one correspondence between the smooth solutions to the Laplace equation inside a sphere and an arbitrary function on the sphere. Indeed, this was proved in \cite{Mirbabayi:2016xvc}. Also it should be noted that since the residual symmetries are defined all-over the spacetime, there is no need to the ``antipodal matching'' used in \cite{He:2014cra} to relate asymptotic symmetries of future and past null infinity.

%%%%%%%%%%%%%%%%%%%%%%%%%%%%%%%%%%%%%%%%%%%%%%%%%%%%%%%

\section*{Acknowledgements}
The author would like to thank specially M.M. Sheikh-Jabbari for many discussions in the course of this project. Also I am grateful to H. Afshar, S. Avery, G. Comp\`ere, J. Evslin, G. Giribet, T. He, M. Mirbabayi, M. Pate and D. Van den Bleeken for their comments on the paper. I also appreciate the organizers of the workshop on quantum aspects of black holes in Yerevan during August 2016. The author would like to thank Bonyad Melli Nokhbegan (BMN) and SarAmadan club of Iran for the partial support.

\appendix
\section{Algebra of charges in covariant phase space}\label{appendix-A-sec}
Here we briefly discuss the algebra of charges in the covariant phase space. Defining a coordinate system over the infinite dimensional phase space, we can write 
\begin{align}\label{Omega_AB def}
\int_\Sigma \bomega(\psi,\de_{1}\psi,\de_{2}\psi)&=\Omega_{AB}\,[\de_{1}\psi]^A\,[\de_{2}\psi]^B\,.
\end{align}
The right hand side is the symplectic two form contracted with two vectors $[\de_{\l_i}\psi]^A$ tangent to the phase space. Also equation \eqref{var H def} is translated to 
\begin{align}\label{var H 2}
\p_A H_\l&=\Omega_{AB}[\de_{\l}\psi]^B\,.
\end{align}
The inverse of the symplectic form $\Omega_{AB}$ defines a Poisson bracket between functions over the phase space through
\begin{align}
\{F,G\}&=\Omega^{AB}\,\p_A F\, \p_B G\,.
\end{align}
Fortunately the Poisson bracket of charges can be computed without knowing the explicit form of the inverse. This is because 
\begin{align}
\{H_{\l_1},H_{\l_2}\}&=\Omega^{AB}\p_A H_{\l_1} \p_B H_{\l_2}\,.
\end{align}
Using \eqref{var H 2} and the fact that $\Omega^{AC}\Omega_{CB}=\de^A_{\;B}$, we arrive at
\begin{align}
\{H_{\l_1},H_{\l_2}\}&=\Omega_{AB}[\de_{\l_1}\psi]^A[\de_{\l_2}\psi]^B\,.
\end{align}
Translating back to spacetime notation using \eqref{Omega_AB def}, we conclude that 
\begin{align}
\{H_{\l_1},H_{\l_2}\}&=\int_\Sigma \bomega(\psi,\de_{\l_1}\psi,\de_{\l_2}\psi)
\end{align}

%\bibliography{Biblio}

\begin{thebibliography}{10}

\bibitem{Noether:1918zz}
E.~Noether, ``{Invariant Variation Problems},''
  \href{http://dx.doi.org/10.1080/00411457108231446}{{\em Gott. Nachr.}
  {\bfseries 1918} (1918) 235--257},
  \href{http://arxiv.org/abs/physics/0503066}{{\ttfamily arXiv:physics/0503066
  [physics]}}.
[Transp. Theory Statist. Phys.1,186(1971)].
%%CITATION = PHYSICS/0503066;%%.

\bibitem{Barnich:2001jy}
G.~Barnich and F.~Brandt, ``{Covariant theory of asymptotic symmetries,
  conservation laws and central charges},''
  \href{http://dx.doi.org/10.1016/S0550-3213(02)00251-1}{{\em Nucl. Phys.}
  {\bfseries B633} (2002) 3--82},
\href{http://arxiv.org/abs/hep-th/0111246}{{\ttfamily arXiv:hep-th/0111246
  [hep-th]}}.
%%CITATION = HEP-TH/0111246;%%.

\bibitem{Iyer:1994ys}
V.~Iyer and R.~M. Wald, ``{Some properties of Noether charge and a proposal for
  dynamical black hole entropy},''
  \href{http://dx.doi.org/10.1103/PhysRevD.50.846}{{\em Phys. Rev.} {\bfseries
  D50} (1994) 846--864},
\href{http://arxiv.org/abs/gr-qc/9403028}{{\ttfamily arXiv:gr-qc/9403028
  [gr-qc]}}.
%%CITATION = GR-QC/9403028;%%.

\bibitem{Compere:2009dp}
G.~Compere, K.~Murata, and T.~Nishioka, ``{Central Charges in Extreme Black
  Hole/CFT Correspondence},''
  \href{http://dx.doi.org/10.1088/1126-6708/2009/05/077}{{\em JHEP} {\bfseries
  05} (2009) 077},
\href{http://arxiv.org/abs/0902.1001}{{\ttfamily arXiv:0902.1001 [hep-th]}}.
%%CITATION = ARXIV:0902.1001;%%.

\bibitem{Sheikh-Jabbari:2016lzm}
M.~M. Sheikh-Jabbari, ``{Residual Diffeomorphisms and Symplectic Softs Hairs:
  The Need to Refine Strict Statement of Equivalence Principle},''
\href{http://arxiv.org/abs/1603.07862}{{\ttfamily arXiv:1603.07862 [hep-th]}}.
%%CITATION = ARXIV:1603.07862;%%.

\bibitem{Avery:2015rga}
S.~G. Avery and B.~U.~W. Schwab, ``{Noether's Second Theorem and Ward
  Identities for Gauge Symmetries},''
\href{http://arxiv.org/abs/1510.07038}{{\ttfamily arXiv:1510.07038 [hep-th]}}.
%%CITATION = ARXIV:1510.07038;%%.

\bibitem{Mirbabayi:2016xvc}
M.~Mirbabayi and M.~Simonovi\'c, ``{Weinberg Soft Theorems from Weinberg
  Adiabatic Modes},''
\href{http://arxiv.org/abs/1602.05196}{{\ttfamily arXiv:1602.05196 [hep-th]}}.
%%CITATION = ARXIV:1602.05196;%%.

\bibitem{Compere:2015knw}
G.~Comp\'ere, P.-J. Mao, A.~Seraj, and M.~M. Sheikh-Jabbari, ``{Symplectic and
  Killing symmetries of AdS$_{3}$ gravity: holographic vs boundary
  gravitons},'' \href{http://dx.doi.org/10.1007/JHEP01(2016)080}{{\em JHEP}
  {\bfseries 01} (2016) 080},
\href{http://arxiv.org/abs/1511.06079}{{\ttfamily arXiv:1511.06079 [hep-th]}}.
%%CITATION = ARXIV:1511.06079;%%.

\bibitem{Arnowitt:1962hi}
R.~L. Arnowitt, S.~Deser, and C.~W. Misner, ``{The Dynamics of general
  relativity},'' \href{http://dx.doi.org/10.1007/s10714-008-0661-1}{{\em Gen.
  Rel. Grav.} {\bfseries 40} (2008) 1997--2027},
\href{http://arxiv.org/abs/gr-qc/0405109}{{\ttfamily arXiv:gr-qc/0405109
  [gr-qc]}}.
%%CITATION = GR-QC/0405109;%%.

\bibitem{Brown:1986nw}
J.~D. Brown and M.~Henneaux, ``{Central Charges in the Canonical Realization of
  Asymptotic Symmetries: An Example from Three-Dimensional Gravity},''
\href{http://dx.doi.org/10.1007/BF01211590}{{\em Commun. Math. Phys.}
  {\bfseries 104} (1986) 207--226}.
%%CITATION = CMPHA,104,207;%%.

\bibitem{Henneaux:1992ig}
M.~Henneaux and C.~Teitelboim, {\em {Quantization of gauge systems}}.
\newblock
1992.
\newblock
%%CITATION = INSPIRE-345963;%%.

\bibitem{Blagojevic:2002du}
M.~Blagojevic, {\em {Gravitation and gauge symmetries}}.
\newblock
2002.
\newblock
%%CITATION = INSPIRE-593015;%%.

\bibitem{Lee:1990nz}
J.~Lee and R.~M. Wald, ``{Local symmetries and constraints},''
\href{http://dx.doi.org/10.1063/1.528801}{{\em J. Math. Phys.} {\bfseries 31}
  (1990) 725--743}.
%%CITATION = JMAPA,31,725;%%.

\bibitem{Crnkovic:1986ex}
C.~Crnkovic and E.~Witten, ``Covariant description of canonical formalism in
  geometrical theories, 1986, \textit{Print-86-1309 (Princeton)},''.

\bibitem{Wald:1993nt}
R.~M. Wald, ``{Black hole entropy is the Noether charge},''
  \href{http://dx.doi.org/10.1103/PhysRevD.48.R3427}{{\em Phys. Rev.}
  {\bfseries D48} (1993) 3427--3431},
\href{http://arxiv.org/abs/gr-qc/9307038}{{\ttfamily arXiv:gr-qc/9307038
  [gr-qc]}}.
%%CITATION = GR-QC/9307038;%%.

\bibitem{Compere:2007az}
G.~Comp\'ere, {\em {Symmetries and conservation laws in Lagrangian gauge
  theories with applications to the mechanics of black holes and to gravity in
  three dimensions}}.
\newblock PhD thesis, Vrije U., Brussels, 2007.
\newblock \href{http://arxiv.org/abs/0708.3153}{{\ttfamily arXiv:0708.3153
  [hep-th]}}.
\newblock
\url{http://inspirehep.net/record/758992/files/arXiv:0708.3153.pdf}.
\newblock
%%CITATION = ARXIV:0708.3153;%%.

\bibitem{Seraj:2016cym}
A.~Seraj, {\em {Conserved charges, surface degrees of freedom, and black hole
  entropy}}.
\newblock PhD thesis, IPM, Tehran, 2016.
\newblock \href{http://arxiv.org/abs/1603.02442}{{\ttfamily arXiv:1603.02442
  [hep-th]}}.
\newblock
\url{http://inspirehep.net/record/1426683/files/arXiv:1603.02442.pdf}.
\newblock
%%CITATION = ARXIV:1603.02442;%%.

\bibitem{Skenderis:2002wp}
K.~Skenderis, ``{Lecture notes on holographic renormalization},''
  \href{http://dx.doi.org/10.1088/0264-9381/19/22/306}{{\em Class. Quant.
  Grav.} {\bfseries 19} (2002) 5849--5876},
\href{http://arxiv.org/abs/hep-th/0209067}{{\ttfamily arXiv:hep-th/0209067
  [hep-th]}}.
%%CITATION = HEP-TH/0209067;%%.

\bibitem{Strominger:1997eq}
A.~Strominger, ``{Black hole entropy from near horizon microstates},''
  \href{http://dx.doi.org/10.1088/1126-6708/1998/02/009}{{\em JHEP} {\bfseries
  02} (1998) 009},
\href{http://arxiv.org/abs/hep-th/9712251}{{\ttfamily arXiv:hep-th/9712251
  [hep-th]}}.
%%CITATION = HEP-TH/9712251;%%.

\bibitem{Carlip:1998wz}
S.~Carlip, ``{Black hole entropy from conformal field theory in any
  dimension},'' \href{http://dx.doi.org/10.1103/PhysRevLett.82.2828}{{\em Phys.
  Rev. Lett.} {\bfseries 82} (1999) 2828--2831},
\href{http://arxiv.org/abs/hep-th/9812013}{{\ttfamily arXiv:hep-th/9812013
  [hep-th]}}.
%%CITATION = HEP-TH/9812013;%%.

\bibitem{Afshar:2016uax}
H.~Afshar, D.~Grumiller, and M.~M. Sheikh-Jabbari, ``{Near Horizon Soft Hairs
  as Microstates of Three Dimensional Black Holes},''
\href{http://arxiv.org/abs/1607.00009}{{\ttfamily arXiv:1607.00009 [hep-th]}}.
%%CITATION = ARXIV:1607.00009;%%.

\bibitem{Strominger:2013lka}
A.~Strominger, ``{Asymptotic Symmetries of Yang-Mills Theory},''
  \href{http://dx.doi.org/10.1007/JHEP07(2014)151}{{\em JHEP} {\bfseries 07}
  (2014) 151},
\href{http://arxiv.org/abs/1308.0589}{{\ttfamily arXiv:1308.0589 [hep-th]}}.
%%CITATION = ARXIV:1308.0589;%%.

\bibitem{He:2014cra}
T.~He, P.~Mitra, A.~P. Porfyriadis, and A.~Strominger, ``{New Symmetries of
  Massless QED},'' \href{http://dx.doi.org/10.1007/JHEP10(2014)112}{{\em JHEP}
  {\bfseries 10} (2014) 112},
\href{http://arxiv.org/abs/1407.3789}{{\ttfamily arXiv:1407.3789 [hep-th]}}.
%%CITATION = ARXIV:1407.3789;%%.

\bibitem{He:2014laa}
T.~He, V.~Lysov, P.~Mitra, and A.~Strominger, ``{BMS supertranslations and
  Weinberg's soft graviton theorem},''
  \href{http://dx.doi.org/10.1007/JHEP05(2015)151}{{\em JHEP} {\bfseries 05}
  (2015) 151},
\href{http://arxiv.org/abs/1401.7026}{{\ttfamily arXiv:1401.7026 [hep-th]}}.
%%CITATION = ARXIV:1401.7026;%%.

\bibitem{Kapec:2015ena}
D.~Kapec, M.~Pate, and A.~Strominger, ``{New Symmetries of QED},''
\href{http://arxiv.org/abs/1506.02906}{{\ttfamily arXiv:1506.02906 [hep-th]}}.
%%CITATION = ARXIV:1506.02906;%%.

\bibitem{Balachandran:1991dw}
A.~P. Balachandran, G.~Bimonte, K.~S. Gupta, and A.~Stern, ``{Conformal edge
  currents in Chern-Simons theories},''
  \href{http://dx.doi.org/10.1142/S0217751X92002106}{{\em Int. J. Mod. Phys.}
  {\bfseries A7} (1992) 4655--4670},
\href{http://arxiv.org/abs/hep-th/9110072}{{\ttfamily arXiv:hep-th/9110072
  [hep-th]}}.
%%CITATION = HEP-TH/9110072;%%.

\bibitem{Balachandran:1994up}
A.~P. Balachandran, L.~Chandar, and A.~Momen, ``{Edge states in gravity and
  black hole physics},''
  \href{http://dx.doi.org/10.1016/0550-3213(95)00622-2}{{\em Nucl. Phys.}
  {\bfseries B461} (1996) 581--596},
\href{http://arxiv.org/abs/gr-qc/9412019}{{\ttfamily arXiv:gr-qc/9412019
  [gr-qc]}}.
%%CITATION = GR-QC/9412019;%%.

\bibitem{Jackiw:2015aoc}
R.~Jackiw and S.~Y. Pi, ``{New Setting for Spontaneous Gauge Symmetry
  Breaking?},'' \href{http://dx.doi.org/10.1007/978-3-319-31299-6_8}{{\em
  Fundam. Theor. Phys.} {\bfseries 183} (2016) 159--162},
\href{http://arxiv.org/abs/1511.00994}{{\ttfamily arXiv:1511.00994 [hep-th]}}.
%%CITATION = ARXIV:1511.00994;%%.

\bibitem{Weinberg:2003sw}
S.~Weinberg, ``{Adiabatic modes in cosmology},''
  \href{http://dx.doi.org/10.1103/PhysRevD.67.123504}{{\em Phys. Rev.}
  {\bfseries D67} (2003) 123504},
\href{http://arxiv.org/abs/astro-ph/0302326}{{\ttfamily arXiv:astro-ph/0302326
  [astro-ph]}}.
%%CITATION = ASTRO-PH/0302326;%%.

\bibitem{Brown:1986ed}
J.~D. Brown and M.~Henneaux, ``{On the Poisson Brackets of Differentiable
  Generators in Classical Field Theory},''
\href{http://dx.doi.org/10.1063/1.527249}{{\em J. Math. Phys.} {\bfseries 27}
  (1986) 489--491}.
%%CITATION = JMAPA,27,489;%%.

\bibitem{Compere:2014cna}
G.~Comp\'ere, L.~Donnay, P.-H. Lambert, and W.~Schulgin, ``{Liouville theory
  beyond the cosmological horizon},''
  \href{http://dx.doi.org/10.1007/JHEP03(2015)158}{{\em JHEP} {\bfseries 03}
  (2015) 158},
\href{http://arxiv.org/abs/1411.7873}{{\ttfamily arXiv:1411.7873 [hep-th]}}.
%%CITATION = ARXIV:1411.7873;%%.

\bibitem{Compere:2015bca}
G.~Comp\'ere, K.~Hajian, A.~Seraj, and M.~M. Sheikh-Jabbari, ``{Wiggling Throat
  of Extremal Black Holes},''
  \href{http://dx.doi.org/10.1007/JHEP10(2015)093}{{\em JHEP} {\bfseries 10}
  (2015) 093},
\href{http://arxiv.org/abs/1506.07181}{{\ttfamily arXiv:1506.07181 [hep-th]}}.
%%CITATION = ARXIV:1506.07181;%%.

\bibitem{Compere:2015mza}
G.~Comp\'ere, K.~Hajian, A.~Seraj, and M.~M. Sheikh-Jabbari, ``{Extremal
  Rotating Black Holes in the Near-Horizon Limit: Phase Space and Symmetry
  Algebra},'' \href{http://dx.doi.org/10.1016/j.physletb.2015.08.027}{{\em
  Phys. Lett.} {\bfseries B749} (2015) 443--447},
\href{http://arxiv.org/abs/1503.07861}{{\ttfamily arXiv:1503.07861 [hep-th]}}.
%%CITATION = ARXIV:1503.07861;%%.

\bibitem{Hajian:2015xlp}
K.~Hajian and M.~M. Sheikh-Jabbari, ``{Solution Phase Space and Conserved
  Charges: Charges Associated with Exact Symmetries, A General Formulation},''
\href{http://arxiv.org/abs/1512.05584}{{\ttfamily arXiv:1512.05584 [hep-th]}}.
%%CITATION = ARXIV:1512.05584;%%.

\bibitem{Compere:2016hzt}
G.~Compere and J.~Long, ``{Classical static final state of collapse with
  supertranslation memory},''
  \href{http://dx.doi.org/10.1088/0264-9381/33/19/195001}{{\em Class. Quant.
  Grav.} {\bfseries 33} no.~19, (2016) 195001},
\href{http://arxiv.org/abs/1602.05197}{{\ttfamily arXiv:1602.05197 [gr-qc]}}.
%%CITATION = ARXIV:1602.05197;%%.

\bibitem{Strominger:2015bla}
A.~Strominger, ``{Magnetic Corrections to the Soft Photon Theorem},''
  \href{http://dx.doi.org/10.1103/PhysRevLett.116.031602}{{\em Phys. Rev.
  Lett.} {\bfseries 116} no.~3, (2016) 031602},
\href{http://arxiv.org/abs/1509.00543}{{\ttfamily arXiv:1509.00543 [hep-th]}}.
%%CITATION = ARXIV:1509.00543;%%.

\bibitem{Campiglia:2016efb}
M.~Campiglia and A.~Laddha, ``{Sub-subleading soft gravitons and large
  diffeomorphisms},''
\href{http://arxiv.org/abs/1608.00685}{{\ttfamily arXiv:1608.00685 [gr-qc]}}.
%%CITATION = ARXIV:1608.00685;%%.





\bibitem{Avery:2016zce}
S.~G. Avery and B.~U.~W. Schwab, ``{Soft Black Hole Absorption Rates as
  Conservation Laws},''
\href{http://arxiv.org/abs/1609.04397}{{\ttfamily arXiv:1609.04397 [hep-th]}}.
%%CITATION = ARXIV:1609.04397;%%.

\bibitem{Thorne:1980ru}
K.~S. Thorne, ``{Multipole Expansions of Gravitational Radiation},''
\href{http://dx.doi.org/10.1103/RevModPhys.52.299}{{\em Rev. Mod. Phys.}
  {\bfseries 52} (1980) 299--339}.
%%CITATION = RMPHA,52,299;%%.

\bibitem{Conde:2016csj}
E.~Conde and P.~Mao, ``{Comments on Asymptotic Symmetries and the Sub-leading
  Soft Photon Theorem},''
\href{http://arxiv.org/abs/1605.09731}{{\ttfamily arXiv:1605.09731 [hep-th]}}.
%%CITATION = ARXIV:1605.09731;%%.

\bibitem{bronzan1971magnetic}
J.~Bronzan, ``The magnetic scalar potential,'' {\em American Journal of
  Physics} {\bfseries 39} (1971) 1357--1359.

\end{thebibliography}
\providecommand{\href}[2]{#2}\begingroup\raggedright\endgroup

\end{document}